\documentclass[a4paper,11pt]{article}
\usepackage{amsthm,amsmath,amsfonts,amssymb}

\newcommand{\one}{\mathbf{1}}
\newcommand{\lap}{\Delta}

\newcommand{\C}{\mathbb C}
\newcommand{\R}{\mathbb R}
\newcommand{\N}{\mathbb N}
\newcommand{\Z}{\mathbb Z}

\newcommand{\rkl}{\rangle}
\newcommand{\lkl}{\langle}
\newcommand{\be}{\begin{equation}}
\newcommand{\ee}{\end{equation}}
\newcommand{\beao}{\begin{eqnarray*}}
\newcommand{\eeao}{\end{eqnarray*}}
\newcommand{\nn}{\nonumber}

\newcommand{\D}{\mathcal{D}}

\newcommand{\Hel}{\mathcal{H}_{el}}
\newcommand{\F}{\mathcal{F}}
\newcommand{\cH}{\mathcal{H}}
\newcommand{\cO}{\mathcal{O}}

\newcommand{\cS}{\mathcal{S}}
\newcommand{\hel}{H_{\mathrm{el}}}
\newcommand{\uf}{\underline{f}}

\newcommand{\uh}{\underline{h}}
\newcommand{\Ran}{\mathrm{Ran}}

\newcommand{\supp}{\mathrm{supp}}

\newcommand{\fA}{\mathbf{A}}
\newcommand{\fp}{\mathbf{p}}

\newcommand{\fk}{\mathbf{k}}
\newcommand{\fx}{\mathbf{x}}
\newcommand{\fG}{\mathbf{G}}

\newcommand{\ad}{\mathrm{ad}}

\newcommand{\ol}{\overline}
\newcommand{\feps}{\mbox{\boldmath $\varepsilon$}}
\newcommand{\eps}{\varepsilon}

\newcommand{\sprod}[2]{\mbox{$\langle #1,#2 \rangle$}}       

\newtheorem{theorem}{Theorem}[section]
\newtheorem{lemma}[theorem]{Lemma}

\newtheorem{proposition}[theorem]{Proposition}
\theoremstyle{plain}
\begin{document}

\title{Asymptotic Electromagnetic
Fields in Non-relativistic QED: the Problem of Existence Revisited}

\author{\vspace{5pt} Marcel Griesemer$^1$ and Heribert Zenk$^2$\\
\vspace{-4pt}\small{$1.$ Fachbereich Mathematik, Universit\"at
Stuttgart,} \\
\small{D--70569 Stuttgart, Germany}\\
\vspace{-4pt}\small{$2.$ Mathematisches Institut,
Ludwig-Maximilians-Universit\"at
M\"unchen,} \\
\small{D-80333 M\"unchen, Germany}}
\date{}
\maketitle
\begin{abstract}
This paper is devoted to the scattering of photons at electrons in
models of non-relativistic quantum mechanical particles coupled
minimally to the soft modes of the quantized electromagnetic
field. We prove existence of scattering states involving an
arbitrary number of asymptotic photons of arbitrarily high energy.
Previously, upper bounds on the photon energies seemed necessary
in the case of $n>1$ asymptotic photons and non-confined,
non-relativistic charged particles.
%
\end{abstract}
%
%
\section{Introduction} \label{sec1} \setcounter{equation}{0}

In this paper we study the scattering of electromagnetic fields at
charged particles in the standard (or Pauli-Fierz) model of
non-relativistic quantum electrodynamics. The first problem to be
addressed in this context concerns the \emph{existence of
asymptotic electromagnetic fields}. In the case where the
asymptotic radiation consists of one photon only, there is a
simple solution to this problem \cite{FGS1}: first a propagation estimate is
used to turn an upper bound on the energy distribution of the
charged particles into an upper bound on their asymptotic propagations
speeds. Propagation speeds strictly below the speed of
light are achieved in \emph{non-relativistic} models with an energy bound that is sufficiently low.
In relativistic models, any finite energy bound is sufficient. By Huygens's principle, the strength of
interaction of a freely propagating photon and charged particles
below the speed of light decays at an integrable rate. Hence, by
Cook's argument, the proof is complete. This paper is concerned
with the case of non-relativistic particles and the existence of
electromagnetic fields consisting of $n\geq 1$ photons. This problem
can be reduced to the case $n=1$ by imposing
a suitable bound on the energy of the asymptotic radiation \cite{FGS1}.
We show that such a bound is not necessary: the one-photon
result from \cite{FGS1} generalized readily to an arbitrary number
of asymptotic photons and so do the key elements of its proof. The
main result of this paper is Theorem~\ref{thm1-1}, below. It will be
used in a forthcoming analysis of photo-ionization and it allows
one to simplify the definitions of the scattering operators for
Rayleigh and Compton scattering \cite{FGS2, FGS3, FGS4}.

Note that the phenomenon of massive particles moving faster than
the speed of light, which is at the heart of the problem solved
in this paper, \emph{does} occur in (pseudo-) relativistic
models describing massive particles inside a space-filling
background material with index of refraction larger than one. It
is not merely an artefact of non-relativistic models.

To describe our result in mathematical terms, we confine ourselves
to a one-electron system and we neglect the spin of the electron.
By the methods to be described one can equally handle systems of
arbitrary (finite) numbers of charged particles from several
species. The Hilbert-space $\cH$ of our system is thus the tensor
product $L^2(\R^3)\otimes\F$ where
$$
         \F:=\oplus_{n\geq 0}
         \big[S_n\otimes^{n}L^2(\R^3\times\{1,2\})\big]
$$
denotes the symmetric Fock space over $L^2(\R^3\times\{1,2\})$,
the space of transversal photons. Here $S_n$ denoted the
projection on $L^2(\R^3\times\{1,2\})^n$ onto the subspace of the
symmetric functions of
$(\fk_1,\lambda_1),\ldots,(\fk_n,\lambda_n)\in\R^3\times\{1,2\}$.
Let $N_f$ denote the number operator in $\F$ and let $a(h)$ and
$a^{*}(h)$ be the usual annihilation and creation operators
associated with a function $h\in L^2(\R^3\times\{1,2\})$. That is,
for $\Psi\in D(N_f^{1/2})$,
$$
     [a^{*}(h)\Psi]^{(n)} = \sqrt{n}S_n (h\otimes\Psi^{(n-1)}),
$$
where $\Psi^{(n)}$ denotes the $n$-photon component
of $\Psi$. The annihilation operator $a(h)$ is the adjoint of
$a^{*}(h)$. For the Hamiltonian of the system we choose
\begin{equation}\label{ham}
   H_{}=(\fp+\alpha^{\frac{3}{2}} \fA(\alpha \fx))^2 + V + H_f,
\end{equation}
where $H_f$ denotes the field energy operator, which is the second
quantization of the operator in $L^2(\R^3\times\{1,2\})$ defined
by multiplication with $\omega(\fk)=|\fk|$, and $\fA(\alpha\fx)$ is the
UV-cutoff quantized vector potential in Coulomb gauge, that is,
$$
    \fA(\alpha\fx) = a(\fG_{\fx})+ a^{*}(\fG_{\fx}), \qquad
    \fG_{\fx}(\fk,\lambda)
    := \frac{\kappa(\fk)}{\sqrt{2|\fk|}}
   \feps(\fk,\lambda)e^{-i\alpha\fk\cdot\fx},
$$
where $\feps(\fk,\lambda)\in\R^3$, $\lambda=1,2$, are orthonormal
polarization vectors perpendicular to $\fk$ and
$\kappa\in\cS(\R^3)$ is an ultraviolet cutoff chosen from the
space $\cS(\R^3)$ of rapidly decreasing functions. The operator
$V$ is a multiplication operator  with a real-valued function from
$L^2_{\rm loc}(\R^3)$ denoted by $V$ as well. We assume that $V$
is infinitesimally operator bounded with respect to the Laplacian
$\Delta$, which is satisfied by the Coulomb potentials of all
atoms and molecules. The Hamiltonian $H$ is self-adjoint on the
domain of $-\Delta + H_f$ and essentially self-adjoint on any core
of this operator \cite{Hiroshima2002, HH2008}. We have chosen
atomic units where $\hbar$, the speed of light $c$, and
$2m\alpha^2$, which is four times the Rydberg-energy, are equal to
one. Here and in \eqref{ham} $\alpha$ denotes the fine structure
constant, which is equal to half of the Bohr radius in our units.

The main purpose of this paper is to establish existence of
scattering states of the form
\begin{equation}\label{intro1}
    a_{+}^{*}(h_1)\cdots a_{+}^{*}(h_n)\Psi, \qquad h_i\in L^2(\R^3\times\{1,2\})
\end{equation}
or $a_{-}^{*}(h_1)\cdots a_{-}^{*}(h_n)\Psi$ where the asymptotic
creation operators $a_{\pm}^{*}(h_i)$ are given by
\begin{equation}\label{intro2}
   a_{\pm}^{*}(h)\Psi = \lim_{t\to\pm\infty}
   e^{iHt}a^{*}(h_{i,t})e^{-iHt}\Psi,\qquad h_{i,t}:=e^{-i\omega
   t}h_i,
\end{equation}
and defined on the space of vectors $\Psi\in D(|H|^{1/2})$ for
which the limit \eqref{intro2} exists. Formally it is clear that
\begin{eqnarray}
    \lefteqn{e^{-iHt}a_{\pm}^{*}(h_1)\cdots
    a_{\pm}^{*}(h_n)\Psi}\nonumber\\
    &=& a^{*}(h_{1,t})\cdots a^{*}(h_{n,t})e^{-iHt}\Psi
    + o(1),\qquad (t\to\pm\infty).\label{intro3}
\end{eqnarray}
Hence the vector \eqref{intro1} describes a state containing $n$
photons with given wave functions $h_1,\ldots,h_n$ whose dynamics
is asymptotically free in the distant future in the sense of
equation \eqref{intro3}. An important subspace of $\cH$ which
belongs to the domain of all asymptotic field operators is the
space of bound states, $\cup_{\lambda<\Sigma}\cH_{\lambda}$, where
$\cH_{\lambda}= \Ran\one_{(-\infty,\lambda]}(H)$ is the spectral
subspace of $H$ associated with the interval $(-\infty,\lambda]$,
and $\Sigma$ is the \emph{ionization threshold} of the Hamiltonian
$H$:
\begin{equation}
   \Sigma = \lim_{R\to\infty}\left(\inf_{\varphi\in
   D_R,\, \|\varphi\|=1}\sprod{\varphi}{H \varphi}\right),
\end{equation}
where $D_R:=\{\varphi\in D(H)| \chi(|x|\leq R)\varphi=0\}$. In a
state $\Psi\in \cH_{\lambda}$ the electron is exponentially
localized in the sense that $e^{\eps|x|}\Psi\in \cH$ for $\eps$
sufficiently small \cite{Gri2004}. The following theorem is our main
result.

\begin{theorem} \label{thm1-1}
Let $E < \Sigma + \frac{1}{4\alpha^2}$, $N \in \N$ and $h_1,...,h_N \in
L^2(\R^3 \times \{1,2\})$ with
$$
  \sum_{\lambda=1,2} \int |h_l(\fk,\lambda)|^2
  (|\fk|^2+\frac{1}{|\fk|})d\fk < \infty
$$
for $l=1,...,N$. Then for each $\Psi \in \Ran
\one_{(-\infty,E]}(H)$ the limit
\begin{equation}\label{gl1-16}
  \lim_{t \to \infty} e^{itH} a^{\#}(h_{1,t}) \cdots
   a^{\#}(h_{N,t}) e^{-itH} \Psi
\end{equation}
exists for any given succession of creation operators $a^{\#}=a^*$
and annihilation operators $a^{\#}=a$ and it equals
\begin{equation} \label{gl1-17}
   a^{\#}_+(h_1) \cdots a^{\#}_+(h_N) \Psi.
\end{equation}
An analog result holds for the limit $t \to -\infty$.
\end{theorem}

This theorem shows, in particular, that the domain of an
asymptotic annihilation or creation operator $a^{\#}_{+}(h)$, with
$h,\omega h, \omega^{-1/2}h \in L^2(\R^3\times\{1,2\})$ contains
the span of all vectors of the form \eqref{gl1-17} with
$h_1,...,h_N$ and $\Psi\in\cH$ satisfying the assumptions of
Theorem~\ref{thm1-1}.

Theorem~\ref{thm1-1} is to be compared with Theorem~6 of
\cite{FGS1}. It shows that the bound on the photon energies imposed
there is unnecessary. In the case $N=1$ the statement of
Theorem~\ref{thm1-1} and its proof below reduce to the Theorem
4,~(i) from \cite{FGS1} and the proof given there. Suitable
adjustments of that proof allow us to prove existence of the limit
\eqref{gl1-16} for arbitrary $N\geq 1$. That \eqref{gl1-16} agrees
with the composition of the operators
$a^{\#}_+(h_1),\ldots,a^{\#}_+(h_N)$ applied to $\Psi$ is
established in a second, independent step.

The main ingredients for the proof of \eqref{gl1-16} are a
propagation estimate for the electron and stationary phase
arguments for the evolution of the photon, that is, Huygens'
principle. The condition that the energy distribution of $\Psi$ is
supported below $E<\Sigma+ \frac{1}{4\alpha^2}$
implies that the (kinetic) energy
of an ionized electron described by $\Psi$ is strictly below
$\frac{1}{4\alpha^2}$
which is $mc^2/2$ in our units. Hence the speed of that electron
is strictly below the speed of light. See the introduction of
\cite{FGS1} for detailed explanations of these ideas.

Previous to this paper the existence of asymptotic creation and
annihilation operators was established in \cite{HK1969, Ammari2000} for
massive bosons, in \cite{Arai1983a,Arai1983b} for (massless)
photons in explicitly solvable models from non-relativistic QED, and in
\cite{HubnerSpohn1995b, Gerard2002} for massless bosons in spin-boson models.
In \cite{FGS1} the existence of many-photon scattering
states is established both in non-relativistic, and in
pseudo-relativistic models from QED.

{\emph{ Acknowledgement:}} M.G. thanks Heinz Siedentop and Laszlo
Erd\"os for the hospitality at the University of Munich, where
this paper was finished.

\section{The Proof} \label{sec2}
\setcounter{equation}{0}
We divide the proof of Theorem \ref{thm1-1} into two parts, the
existence of the limit in (\ref{gl1-16}) is established in
Proposition~\ref{pr2-1} and the equality of (\ref{gl1-16}) and
(\ref{gl1-17}) is Proposition \ref{pro2-2}. We begin by
introducing some useful notations. The inner product of two
functions $f,g\in L^2(\R^3 \times \{1,2\})$ is denoted by
$\sprod{f}{g}$, that is,
$$
     \sprod{f}{g}:= \sum_{\lambda=1,2} \, \int
     \overline{f(\fk,\lambda)}g(\fk,\lambda) d\fk.
$$
By $L^2_{\omega}(\R^3 \times \{1,2\})$ we denote the space of
functions $f\in L^2(\R^3 \times \{1,2\})$ with
$$
    \|f\|^2_{\omega} := \sum_{\lambda=1,2}\int|f(\fk,\lambda)|^2
    (1+\omega(\fk)^{-1})\,d\fk
    <\infty.
$$
The assumption on $h_l$ in Theorem~\ref{thm1-1} means that both
$h_l$ and $\omega h_l$ belong to $L^2_{\omega}(\R^3 \times
\{1,2\})$. Note that $L^2_{\omega}(\R^3 \times \{1,2\})$ is
isomorphic to the space $L^2_{T,\omega}$ of square integrable
functions $f:\R^3\to\C^3$ with respect to
$(1+\omega(\fk)^{-1})\,d\fk$, satisfying $\fk\cdot f(\fk)=0$,
almost everywhere. Given a choice of polarization vectors
$\feps(\fk,\lambda)$, $\fk \in \R^3, \lambda \in \{1,2\}$
perpendicular to $\fk$, this isomorphism $\feps: L^2(\R^3 \times
\{1,2\}) \to L^2_{T,\omega}$ is expressed by the equation $(\feps
f)(\fk):=\sum_{\lambda}\feps(\fk,\lambda)f(\fk,\lambda)$.
 If $\uh=(h_1,...,h_N)$ with $h_{l}\in L^2(\R^3 \times
\{1,2\})$ then
$$
    a^{\#}(\uh) := a^{\#}(h_1)\cdots a^{\#}(h_N)
$$
where each factor $a^{\#}(h_l)$ may be either an annihilation
operator or a creation operator on Fock space.


\begin{proposition} \label{pr2-1}
Let $\uh=(h_1,...,h_N) \in [L^2_{\omega}(\R^3 \times \{1,2\})]^N$,
$E < \Sigma + \frac{1}{4\alpha^2}$ and $\Psi=\one_{(-\infty,E]}(H)
\Psi$, then \be a^{\#}_{\pm}(\uh) \Psi:= \lim_{t \to \pm \infty}
e^{itH} a^{\#}(\uh_t) e^{-itH} \Psi \label{gl2-1} \ee exists and
there is a constant $C(N,E)$, such that \be \|a^{\#}_{\pm}(\uh)
\one_{(-\infty,E]}(H) \| \leq C(N,E) \prod_{l=1}^N
\|h_l\|_{\omega}. \label{gl2-2}\ee
\end{proposition}
%
The proof of this Proposition is based on the methods developed in
\cite{FGS1}, and in particular on the propagation estimate \be
\int\limits_1^{\infty} \frac{dt}{t^{\mu}} \|\one_{\{|\fx| \geq
vt\}} e^{-itH} g(H) \Psi \|^2 \leq C \|(1+|\fx|)^{\frac{1}{2}}
g(H)\Psi\|^2, \label{gl2-3} \ee which holds for $\mu > 1/2$ and
$g\in C_0^{\infty}(\R)$ with $\sup\{\lambda \in \R: g(\lambda)
\not=0\} < \Sigma+v^2/4$.

\begin{proof}
We pick $g \in C_0^{\infty}(\R)$ with
$\supp(g)\subset(-\infty,\Sigma+\frac{1}{4\alpha^2})$, $g=1$ on
$(-\infty,E]$, so that $g(H)=1$ on $\Ran\one_{(-\infty,E]}(H)$. By
\eqref{gl2-4} and by part b) of Lemma~\ref{la3}, the operator
$e^{itH} a^{\#}(\uh_t) e^{-itH}g(H)$ is bounded uniformly in
$t\in\R$. Hence it suffices to prove existence of
\begin{equation}\label{gl2-6}
   \lim_{t\to\infty} e^{itH} a^{\#}(\uh_t)e^{-itH}g(H)\Psi
\end{equation}
for vectors $\Psi$ in the dense subspace $\D(\lkl \fx
\rkl^{\frac{1}{2}})$ of $\cH$, where $\lkl \fx \rkl$ denotes the
operator of multiplication with $\lkl \fx \rkl =
(1+\fx^2)^{\frac{1}{2}}$ in $\Hel$. We first prove existence of
the limit \eqref{gl2-6} for $\uh=\uf=(f_1,...,f_N)$ with functions
$f_l$  for which $\feps f_l$ belongs to $C_0^{\infty}(\R^3
\backslash \{0\}, \C^3)$. For notational simplicity, we confine
ourselves to the case, where $a^{\#}(\uf_t)=a^*(f_{1,t}) \cdots
a^*(f_{N,t})$. In the general case $\sprod{\fG_{\fx}}{f_{l,t}}$
needs to be replaced by $-\ol{\lkl \fG_{\fx},f_{l,t} \rkl }$
whenever $a^{\#}(f_{l,t})$ denotes an annihilation operator, which
does not effect our estimates.

By Cook`s argument, the limit of $\Psi(t)=e^{itH}
a^{\#}(\uf_t)e^{-itH}g(H)\Psi$ as $t\to\infty$ exists, provided
that
\begin{equation}
  \int\limits_1^{\infty} \Big\| \frac{d}{dt} \Psi(t)\Big\| dt <
  \infty. \label{gl2-7}
\end{equation}
To prove \eqref{gl2-7}, we choose $\varepsilon >0$ so small, that
$\sup(\supp g) < \Sigma + \frac{1}{4\alpha^2} (1-2\varepsilon)^2$
and we pick $\chi_1, \chi_2 \in C^{\infty}(\R, [0,1])$, such that
$\chi_1+\chi_2=1$, $\chi_1(s)=0$ for $s \leq 1-2\varepsilon$ and
$\chi_1(s)=1$ for $s \geq 1-\varepsilon$. Let
$\chi_{1,t}(\fx):=\chi_1(\alpha|\fx|/t)$ and
$\chi_{2,t}(\fx):=\chi_2(\alpha|\fx|/t)$. Then \label{rechnung}
\begin{eqnarray}
\Psi'(t)&=&ie^{itH} [(\fp+\alpha^{\frac{3}{2}}\fA(\alpha
\fx))^2,a^{*}(\uf_t)]
e^{-itH} g(H) \Psi \nn \\
&=& \sum_{l=1}^N  e^{itH}2i\alpha^{\frac{3}{2}} \lkl \fG_{\fx},
f_{l,t} \rkl a^{*}(f_{1,t}) \cdots a^{*}(f_{l-1,t})
 \label{gl2-8} \\
&& \qquad \cdot(\fp +\alpha^{\frac{3}{2}} \fA(\alpha \fx))
a^{*}(f_{l+1,t}) \cdots a^{*}(f_{N,t}) e^{-itH} g(H) \Psi, \nn
\end{eqnarray}
where the three components of $\lkl \fG_{\fx}, f_{l,t}
\rkl\in\C^3$ are to be considered as multiplication operator in
$\Hel$. Since $\supp(\chi_{2,t}) \subseteq \{\fx \in \R^3:
\frac{\alpha |\fx|}{t} < 1- \varepsilon\}$ and
$|\nabla_{\fk}(i\alpha \fk \cdot \fx-i\omega(\fk)t)| = |\alpha\fx
-t\frac{\fk}{|\fk|}|>|t|\eps$ on this set, it follows, by
stationary phase arguments, that
\begin{equation}
  | \lkl \fG_{\fx}, f_{l,t} \rkl \chi_{2,t}(\fx)| \leq
\frac{c_l}{1+t^2}, \label{gl2-10}
\end{equation}
while for all $\fx \in \R^3$ and all $t \in \R$
\begin{equation}\label{gl2-11}
  | \lkl \fG_{\fx}, f_{l,t} \rkl | \leq \frac{c_l}{1+|t|}
\end{equation}
by Theorem XI.18 in \cite{ReedSimon3}. We write $\lkl \fG_{\fx},f_{l,t}
\rkl = \lkl \fG_{\fx},f_{l,t} \rkl \chi_{1,t} + \lkl
\fG_{\fx},f_{l,t} \rkl \chi_{2,t}$ and estimate the two
contributions to \eqref{gl2-8} separately. By Lemma \ref{la3} and
\eqref{gl2-10} the contribution of
$\sprod{\fG_{\fx}}{f_{l,t}}\chi_{2,t}$ to \eqref{gl2-8} is
integrable with respect to $t \in \R$.
As for the contribution of
$\sprod{\fG_{\fx}}{f_{l,t}}\chi_{1,t}$, due to \eqref{gl2-11} it
is enough to prove integrability with respect to $t \in
[1,\infty)$ of
\begin{eqnarray}
\lefteqn{ \frac{1}{t} \chi_{1,t}
a^{*}(f_{1,t}) \cdots a^{*}(f_{l-1,t}) (\fp +\alpha^{\frac{3}{2}}
\fA(\alpha \fx)) a^{*}(f_{l+1,t}) \cdots a^{*}(f_{N,t})
e^{-itH} g(H) \Psi \nn} \\
&=&\frac{1}{t}
a^{*}(f_{1,t}) \cdots a^{*}(f_{l-1,t})
(\fp +\alpha^{\frac{3}{2}} \fA(\alpha \fx))
a^{*}(f_{l+1,t}) \cdots a^{*}(f_{N,t})  \chi_{1,t}
e^{-itH} g(H) \Psi \nn \\
&&+ \frac{1}{t}(i \nabla \chi_{1,t}) a^{*}(f_{1,t}) \cdots
a^{*}(f_{l-1,t}) a^{*}(f_{l+1,t}) \cdots a^{*}(f_{N,t}) e^{-itH}
g(H) \Psi. \label{gl2-12}
\end{eqnarray}
Since $|\nabla\chi_{1,t}| = \cO(t^{-1})$ the second term of
\eqref{gl2-12} is of order $\cO(t^{-2})$, hence integrable. In the
first term we use
\begin{eqnarray}
\lefteqn{ \chi_{1,t}=(H+i)^{-N} \chi_{1,t}
(H+i)^N -
[(H+i)^{-N}, \chi_{1,t}] (H+i)^N}\nn\\
&&=(H+i)^{-N} \chi_{1,t} (H+i)^N - \sum_{k=1}^N (H+i)^{-k+1}
[(H+i)^{-1},\chi_{1,t}]
(H+i)^k \nn \\
&&=(H+i)^{-N} \chi_{1,t} (H+i)^N + \sum_{k=1}^N (H+i)^{-k}
[H,\chi_{1,t}] (H+i)^{k-1} \label{gl2-13}
\end{eqnarray}
and we claim, that each term in (\ref{gl2-12}) originating from
the sum of commutators in (\ref{gl2-13}) is of order $\cO(t^{-2})$
due to the additional $t^{-1}$ from $[H,\chi_{1,t}]$. Let`s prove
this for the contribution from $\fp$ in $\fp+\alpha^{\frac{3}{2}}
\fA(\alpha \fx)$. To this end we set
$$
  a^*(\uf_{(l),t}):=a^*(f_{1,t}) \cdots a^*(f_{l-1,t}) a^*(f_{l+1,t})
  \cdots a^*(f_{N,t})
$$
and $g_k(H):=(H+i)^{k-1} g(H)$. By the Cauchy-Schwarz inequality
\begin{gather}
\|a^*(\uf_{(l),t}) \fp (H+i)^{-k}[H,\chi_{1,t}] g_k(H)
\|^2 \leq \|\fp^2 (H+i)^{-k} [H,\chi_{1,t}] g_k(H)\|\nn\\
 \|a(\uf_{(l),t}) a^*(\uf_{(l),t}) (H+i)^{-k} [H,\chi_{1,t}]
g_k(H)\|.\label{gl2-14}
\end{gather}
Since
\begin{equation}
  [H,\chi_{1,t}]=(-2i\nabla \chi_{1,t}) (\fp+\alpha^{\frac{3}{2}}
  \fA(\alpha \fx))-\lap \chi_{1,t} \label{gl2-13a}
\end{equation}
the first factor of \eqref{gl2-14} is bounded by
\begin{gather*}
\|\fp^2 (H+i)^{-1}\| \Big( 2 \|\nabla \chi_{1,t}\| \,
\|(\fp+\alpha^{\frac{3}{2}} \fA(\alpha \fx))(H+i)^{-1}\|\,
\|g_{k+1}(H)\|\\
+ |\lap \chi_{1,t}| \, \|g_k(H)\|\Big) = \cO(t^{-1})
\end{gather*}
and the second factor is bounded by \be C\|(H_f+1)^N
[H,\chi_{1,t}] g_k(H) \| \label{gl2-15} \ee thanks to
(\ref{gl2-4}) and Lemma \ref{la1}. Equation~\eqref{gl2-13a} and
the Cauchy-Schwarz inequality yield:
\begin{eqnarray}
\lefteqn{\|(H_f+1)^N [H,\chi_{1,t}] g_k(H) \|\leq
\frac{C}{t}\Big(\alpha^{\frac{3}{2}}
\|(H_f+1)^N \fA(\alpha \fx) g_k(H)\|\nn} \\
&&+\|\fp^2 g_k(H) \|^{\frac{1}{2}} \|(H_f+1)^{2N}g_k(H)
\|^{\frac{1}{2}}+ \frac{1}{t} \|(H_f+1)^N g_k(H) \|\Big),
\end{eqnarray}
which is again $\cO(t^{-1})$.

So far, we have shown that
\begin{eqnarray}
\lefteqn{ \frac{1}{t} \chi_{1,t}
a^{*}(f_{1,t}) \cdots a^{*}(f_{l-1,t}) (\fp +\alpha^{\frac{3}{2}}
\fA(\alpha \fx)) a^{*}(f_{l+1,t}) \cdots a^{*}(f_{N,t})
e^{-itH} g(H) \Psi \nn}\\
&=&
\Big[ a^{*}(f_{1,t}) \cdots a^{*}(f_{l-1,t})
(\fp +\alpha^{\frac{3}{2}} \fA(\alpha \fx))
a^{*}(f_{l+1,t}) \cdots a^{*}(f_{N,t})
(H+i)^{-N} \Big] \nn\\
&&\frac{1}{t} \chi_{1,t} e^{-itH} F(H) \Psi + \cO(\frac{1}{t^2}),
\label{gl2-17}
\end{eqnarray}
where $F(x)=(x+i)^N g(x)$. By \eqref{glA3a}, the norm of the
operator in brackets is bounded uniformly in time. For the norm of
the vector this operator is applied to, we have
\begin{eqnarray}
\lefteqn{ \int\limits_1^{\infty} \frac{dt}{t} \| \chi_{1,t}
e^{-itH} F(H) \Psi \|
\label{gl3-38}} \\
&\leq& \Bigg[\int\limits_1^{\infty} dt\, t^{-\frac{5}{4}}
\Bigg]^{\frac{1}{2}} \Bigg[ \int\limits_1^{\infty} dt
\,t^{-\frac{3}{4}} \|\one_{\{|\fx| \geq
\frac{|t|}{\alpha}(1-\varepsilon)\}} e^{-itH} F(H)
\Psi \|^2 \Bigg]^{\frac{1}{2}} \nn \\
&\leq& 2 \sqrt{C} \|(1+|\fx|)^{\frac{1}{2}} F(H) \Psi \|. \nn
\end{eqnarray}
by the propagation estimate \eqref{gl2-3} with $\mu=\frac{3}{4}$.
The norm $\|(1+|\fx|)^{\frac{1}{2}} F(H) \Psi \|$ is finite,
because $F(H) \D(\lkl \fx \rkl^{\frac{1}{2}}) \subseteq \D(\lkl
\fx \rkl^{\frac{1}{2}})$ by Lemma~20 of \cite{FGS1}. This
concludes the proof of Proposition~\ref{pr2-1} in the case where
$h_j=f_j$ and $\feps f_j$ belongs to $C_0^{\infty}(\R^3 \backslash
\{0\}, \C^3)$. For the proof in the general case, where $h_j\in
L^2_{\omega}(\R^3\times \{1,2\})$, we use that $C_0^{\infty}(\R^3
\backslash \{0\})$-functions are dense in $L^2_{T,\omega}$, which
follows from the fact, that the projection $\varphi(\fk) \mapsto
\varphi(\fk)-\frac{\fk}{\|\fk\|^2} \lkl \varphi(\fk), \fk \rkl$ of
a vector $\varphi(\fk)$ onto the component perpendicular to $\fk$
leaves $C_0^{\infty}(\R^3 \backslash \{0\})$ invariant. Hence for
given $\varepsilon >0$ there exist functions $f_j \in
L^2_{\omega}(\R^3 \times \{1,2\})$, such that $\feps f_j \in
C_0^{\infty}(\R^3 \backslash \{0\}, \C^3)$ and
$\|f_j-h_j\|_{\omega}<\varepsilon$. Using
\[a^{*}(\uh_t)-a^{*}(\uf_t) =
\sum_{l=1}^N a^{*}(h_{1,t}) \cdots a^{*}(h_{l-1,t})
a^{*}(h_{l,t}-f_{l,t}) a^{*}(f_{l+1,t}) \cdots a^{*}(f_{N,t}) \]
and Lemma \ref{la3} we obtain
\begin{eqnarray}
\lefteqn{\sup_{t
\in \R} \|e^{itH}(a^{*}(\uh_t)-a^{*}(\uf_t)) e^{-itH}
g(H)\Psi\| }\\
&\leq& C_N \sum_{n=1}^N \|h_1\|_{\omega} \cdots
\|h_{l-1}\|_{\omega} \|h_l-f_l\|_{\omega} \|h_{l+1}\|_{\omega}
\cdots \|h_N\|_{\omega} \leq C \varepsilon. \nn
\end{eqnarray}
Hence existence of the limit $a^*_+(\uf)g(H) \Psi$ implies, that
the limit $a^*_+(\uh) g(H) \Psi$ exists as well, and the bound
(\ref{gl2-2}), valid for $\uf$, extends to $\uh \in
[L^2_{\omega}(\R^3 \times \{1,2\})]^N$.
\end{proof}
\noindent The following Lemma generalizes the well-known identity
$[iH,a^{\#}_{\pm}(h)] = a^*_{\pm}(i\omega h)$
to the asymptotic $N$-photon annihilation and creation operators
$a^{\#}_{\pm}(\uh)$
defined by Proposition~\ref{pr2-1}.

\begin{lemma} \label{l2-2}
Suppose that $E<\Sigma+\frac{1}{4\alpha^2}$. Then for all $\uh \in
[L^2_{\omega}(\R^3 \times \{1,2\})]^N$ and all $t \in \R$
\begin{equation}\label{gl3-35b}
  e^{-itH} a^{\#}_{\pm}(\uh) e^{itH} =
  a^{\#}_{\pm}(\uh_{t})
\end{equation}
on $\Ran\one_{(-\infty,E]}(H)$. If $\uh$ and $\omega_l
\uh:=(h_1,...,h_{l-1}, \omega h_l, h_{l+1},...,h_N)$ belong to
$L^2_{\omega}(\R^3 \times \{1,2\})^N$, then
$a^{\#}_{\pm}(\uh)\Ran\one_{(-\infty,E]}(H)\subset D(H)$ and
\begin{equation}\label{gl3-35a}
  \left[iH,a^{\#}_{\pm}(\uh)\right] = \sum_{l=1}^N a^{\#}_{\pm}(i\omega_l \uh)
\end{equation}
on $\Ran\one_{(-\infty,E]}(H)$.
\end{lemma}

\begin{proof}
Equation \eqref{gl3-35b} is obvious from the definition of
$a^{\#}_{\pm}(\uh)$. Now let $\Phi\in D(H)$ and suppose that
$\Psi=\one_{(-\infty,E]}(H)\Psi$. By \eqref{gl3-35b},
\begin{equation}\label{gl3-49}
   \sprod{e^{iHt}\Phi}{a^{\#}_{\pm}(\uh)e^{iHt}\Psi}
   = \sprod{\Phi}{a^{\#}_{\pm}(\uh_{t})\Psi}
\end{equation}
for all $t\in\R$ and we would like to differentiate both sides with
respect to $t$. The left hand side is differentiable because
$a^{\#}_{\pm}(\uh)\one_{(-\infty,E]}(H)$ is a bounded operator and
because $e^{iHt}\Phi$ and $e^{iHt}\Psi$ are differentiable. Hence the
right hand side, $t\mapsto\sprod{\Phi}{a^{\#}_{\pm}(\uh_{t})\Psi}$,
must be differentiable as well. To compute its derivative, we use that
\begin{equation}\label{gl3-50}
   \left\|\frac{1}{\eps}\left(h_{l,\eps}-h_{l}\right) + i\omega
   h_{l}\right\|_{\omega}
   \to 0,\qquad (\eps\to 0),
\end{equation}
as well as \eqref{gl2-2}. Statement \eqref{gl3-50} follows from the assumption on $h_l$,
which implies that both $(1+\omega^{-1})^{1/2}h_l$ and
$(1+\omega^{-1})^{1/2}\omega h_l$ belong to
$L^2(\R^3\times\{1,2\})$. We conclude that
\begin{equation}
   \sprod{iH\Phi}{a^{\#}_{\pm}(\uh)\Psi}+\sprod{\Phi}{a^{\#}_{\pm}(\uh)iH\Psi}
   = -\Big\langle\Phi,\sum_{l=1}^N a^{\#}_{\pm}(i\omega_l \uh)\Psi\Big\rangle.
\end{equation}
Since $H=H^{*}$, it follows that $a^{\#}_{\pm}(\uh)\Psi\in D(H)$, and
that \eqref{gl3-35a} holds.
\end{proof}
\noindent
The following Proposition shows, that (\ref{gl1-16}) and (\ref{gl1-17})
are equal and hence concludes the proof of Theorem \ref{thm1-1}.
\begin{proposition} \label{pro2-2}
Suppose that $h_l,\omega h_l\in L^2_{\omega}(\R^3 \times \{1,2\})$
for $l=1,\ldots,N$, and let $\uh=(h_1,...,h_N)$. If $E < \Sigma +
\frac{1}{4\alpha^2}$ and $\Psi= \one_{(-\infty,E]}(H) \Psi$, then
\be a^{\#}_{\pm} (\uh) \Psi = a^{\#}_{\pm}(h_1) \cdots
a^{\#}_{\pm}(h_N) \Psi, \label{gl2-25}\ee where
$a^{\#}_{\pm}(h_j)$, depending on $j$ may be a creation or an
annihilation operator.
\end{proposition}
\begin{proof}
We show that
\begin{equation}\label{gl3-57}
    a^{*}_{\pm}(\uh) \Psi = a^{*}_{\pm}(h_1)a^{*}_{\pm}(\uh^{(1)})\Psi
\end{equation}
where $\uh^{(1)}:= (h_2,...,h_N)$. Then the proposition follows by
induction in $N$.

From $a^{*}(\uh_t)=a^{*}(h_{1,t})a^{*}(\uh^{(1)}_t)$ it follows
that
\begin{eqnarray*}
\lefteqn{a^{*}_{\pm}(\uh) \Psi - e^{itH} a^{*}(h_{1,t}) e^{-itH}
a^{*}_{\pm}(\uh^{(1)}) \Psi}\\
& =& a^{*}_{\pm} (\uh) \Psi -
e^{itH} a^{*}(\uh_t) e^{-itH} \Psi \\
&& +  e^{itH} a^{*}(\uh_{1,t}) e^{-itH} \Big( e^{itH}
a^{*}(\uh^{(1)}_t) e^{-itH} \Psi - a^{*}_{\pm} (\uh^{(1)}) \Psi
\Big)
\end{eqnarray*}
where the first two term on the right hand side cancel each other in the
limits $t\to\pm\infty$ by Proposition~\ref{pr2-1}. In the third
term we insert $1=(H+i)^{-1}(H+i)$. Since the norm of
$a^{*}(h_{1,t})(H+i)^{-1}$ is bounded uniformly in $t\in\R$, it
remains to estimate the norm of
\begin{eqnarray*}
\lefteqn{(H+i)\Big( e^{itH} a^{*}(\uh^{(1)}_t) e^{-itH} \Psi -
a^{*}_{\pm}
(\uh^{(1)}) \Psi \Big)}\\
 &=& e^{itH} a^{*}(\uh^{(1)}_t) e^{-itH}
(H+i) \Psi - a^{*}_{\pm} (\uh^{(1)}) (H+i) \Psi\\
 &&+\Big[ H,e^{itH} a^{*}(\uh^{(1)}_t) e^{-itH}- a^{*}_{\pm}
(\uh^{(1)}) \Big] \Psi.
\end{eqnarray*}
Again, in the limits $t\to\pm\infty$, the first two terms cancel
each other by Proposition~\ref{pr2-1} and because
$(H+i)\Psi\in\Ran\one_{(-\infty,E]}(H)$. Using \eqref{gl3-35a} to
evaluate the commutator we obtain
\begin{eqnarray*}
\lefteqn{ \Big[H,e^{itH} a^{*}(\uh^{(1)}_t) e^{-itH}-
a^{*}_{\pm} (\uh^{(1)}) \Big] \Psi \nn}\\
&=& \sum_{l=2}^N e^{itH} 2\alpha^{\frac{3}{2}}\lkl \fG_{\fx},
h_{l,t} \rkl a^{*}(h_{2,t}) \cdots a^{*}(h_{l-1,t})\\
&& \qquad\qquad \cdot\big(\fp +\alpha^{\frac{3}{2}} \fA(\alpha
\fx)\big) a^{*}(h_{l+1,t})\cdots a^{*}(h_{N,t})
e^{-itH} \Psi\\
&&+\sum_{l=2}^N  \Big(e^{itH}  a^{*}(\omega_l \uh_t^{(1)})
e^{-itH} \Psi - a^{*}_{\pm} (\omega_l \uh^{(1)}) \Psi \Big).
\end{eqnarray*}
We claim that all terms of these two sums vanish in the limits
$t\to\pm\infty$. For the terms of the second sum this follows from
Proposition~\ref{pr2-1} thanks to the assumption $\omega_l \uh \in
[L^2_{\omega}(\R^3 \times \{1,2\})]^N$. The terms from the first
sum contain a factor $\sprod{\fG_{\fx}}{h_{l,t}}$, where
\begin{equation}\label{gl3-60}
     \sup_{x\in\R^3}\big|\sprod{\fG_{\fx}}{h_{l,t}}\big| \to
     0,\qquad(t\to\infty).
\end{equation}
This is clear from \eqref{gl2-11} in the case where
$\sum_{\lambda}\feps(\fk,\lambda)h_l(\fk,\lambda)$ belongs to
$C_0^{\infty}(\R^3 \backslash \{0\})$, and from there this result extends to
all $h_l$ by the
usual approximation argument. From \eqref{gl3-60} and estimates
similar to those used in the proof of Proposition~\ref{pr2-1}, we
see that the terms of the first sum vanish as well, as
$t\to\pm\infty$. This establishes Equation~\eqref{gl3-57} which
concludes the proof.
\end{proof}


\begin{appendix}
\section{Operator bounds}
\setcounter{equation}{0} In this appendix we collect estimates on
operator norms that are used in the proofs of this paper.

\begin{lemma} \label{la3}
\text{}
\begin{enumerate}
\item[a)] For every $\alpha \in \R$, the operator $\fp^2 (H+i)^{-1}$ is
bounded.
\item[b)] For every $\alpha$ and every $n \in \N$ the operator
$H_f^n (H+i)^{-n} $ is bounded.
\item[c)]
For every $N\in\N$ there is a constant $C_N$, such that for all
$h_1,...,h_N \in L^2_{\omega}(\R^3 \times \{1,2\})$ and all $l \in
\{1,...,N\}$
\begin{eqnarray}
  \|a^*(\uh_t) (H_f+1)^{-\frac{N}{2}} \| &\leq &
C_N \prod_{l=1}^N \|h_l\|_{\omega},\label{gl2-4}\\
\Big\|a^*(h_{1,t}) \cdots a^*(h_{l-1,t})
(\fp +\alpha^{\frac{3}{2}} \fA(\alpha \fx)) &&\nn \\[-0.5cm]
\hspace{1cm} a^*(h_{l+1,t}) \cdots a^*(h_{N,t})
 (H+i)^{-N} \Big\| &\leq& C_N \prod_{\substack{m=1\\m\neq l}}^N\|h_m\|_{\omega} \label{glA3a}
\end{eqnarray}
\end{enumerate}
\end{lemma}

\begin{proof}
By assumption on $V$, $\D(\hel)=\D(\fp^2)$, hence $\fp^2
(\hel+i)^{-1}$ is bounded. Since $\D(H_0)=\D(H)$, see \cite{HH2008},
it follows, that
\[\fp^2 (H+i)^{-1} = \fp^2 (\hel+i)^{-1} (\hel+i) (H_0+i)^{-1}
(H_0+i) (H+i)^{-1} \] is bounded. Part (b) is Lemma~5 in
\cite{FGS1}, and bound \eqref{gl2-4} is the statement of Lemma~17
in that paper. Bound \eqref{glA3a} for the contribution from
$A(\alpha\fx)$ follows from \eqref{gl2-4}, point-wise in
$\fx\in\R^3$. As for the contribution from $\fp$, we note that
\begin{eqnarray*}
   \|a^*(\uh_{(l),t}) \fp (H+i)^{-N} \Psi \|^2 &\leq &
   \|\fp^2 (H+i)^{-N} \Psi \|
   \|a(\uh_{(l),t}) a^*(\uh_{(l),t}) (H_f+1)^{-N}\| \\
&&\|(H_f+1)^N (H+i)^{-N} \Psi\|,
\end{eqnarray*}
by the
Cauchy-Schwarz inequality. The factors on the right hand side are
finite by \eqref{gl2-4} and the parts (a) and (b) that we wave
just established.
\end{proof}
\begin{lemma}\label{la1}
For all $m, n \in \N$ the operator
\begin{equation}
 (H_f+1)^n (H+i)^{-m} (H_f+1)^{-n} \label{glA-3}
\end{equation}
is bounded.
\end{lemma}

\begin{proof}
Let $R:=(H+i)^{-1}$ and $\Phi(h)=a(h)+a^{*}(h)$ in this proof,
where $h\in L^2(\R^3\times\{1,2\})$. Since
\[(H_f+1)^n R^m (H_f+1)^{-n}=((H_f+1)^n R (H_f+1)^{-n})^m \]
it suffices to prove boundedness of (\ref{glA-3}) for $m=1$, which
is equivalent to showing that $[(H_f+1)^n,R] (H_f+1)^{-n}$ is
bounded. We recall from \cite{FGS1}, Appendix B, that
\begin{equation*}
  [(H_f+1)^n, R](H_f+1)^{-n}=\sum_{l=1}^n \binom{n}{l}
  \ad_{H_f}^l(R)(H_f+1)^{-l},
\end{equation*}
where $\ad_{H_f}^0(R)=R$ and
$\ad_{H_f}^{n+1}(R)=[H_f,\ad_{H_f}^n(R)]$. We claim that
$\ad_{H_f}^{l}(R)$ is a bounded operator for all $l\in\N$. To
prove this we note that $\fA(\fx)=\Phi(\fG_{\fx})$ and we define
$$
   W_0 := H-H_0=2\alpha^{\frac{3}{2}} \fp \cdot
\Phi(\fG_{\fx}) + \alpha^3 \Phi(\fG_{\fx})^2
$$
and
\begin{align}
   W_l := \ad_{H_f}^l(W_0)=\ &
   2\alpha^{\frac{3}{2}} (-i)^l \fp \cdot \Phi(i^l\omega^l
   \fG_{\fx}) \nn \\
   &+\alpha^3 \sum_{k=0}^l \binom{l}{k} (-i)^l
   \Phi(i^k\omega^k \fG_{\fx}) \Phi(i^{l-k} \omega^{l-k} \fG_{\fx}).
   \label{glA6}
\end{align}
From $[H_f,R]=-R W_{1}R$ and $[H_f,W_{j}]=W_{j+1}$ we obtain, by
induction in $l$, that
\begin{equation}
\ad_{H_f}^l(R) = \sum_{\substack{j_1,...,j_k=1\\1\leq k\leq l}}^{l}
c_{j_1,...,j_k} RW_{j_1}R \cdots W_{j_k}R\label{glA5}
\end{equation}
with combinatorial factors $c_{j_1,...,j_k} \in \Z$. By
\eqref{glA6} and Lemma~\ref{la3} the operators
$W_{j_1}R,\ldots,W_{j_k}R$ are bounded. Hence \eqref{glA5} shows
that $\ad_{H_f}^l(R)$ is bounded for all $l \in \N$.
\end{proof}
\end{appendix}


\end{document}